\documentclass[
 reprint,
superscriptaddress,
 amsmath,amssymb,
 aps,pre,hyphens,floatfix
]{revtex4-1}
\usepackage{natbib}
\usepackage[]{amsmath}
\usepackage{enumerate}
\usepackage[usenames,dvipsnames,svgnames,table]{xcolor}
\usepackage{IEEEtrantools}
\usepackage{graphicx}
\usepackage{array}
\usepackage{multirow}
\usepackage{verbatim}
\usepackage{float}

\usepackage{changes}
\newcommand{\tcb}[1]{{ #1}}
\newcommand{\stkout}[1]{\ifmmode\text{\sout{\ensuremath{#1}}}\else\sout{#1}\fi}

\begin{document}

\preprint{APS/123-QED}

\title{The behavior of rich-club coefficient in scale-free networks}

\author{Zhihang Liu}
\affiliation{Key Laboratory of Quark and Lepton Physics (MOE) and Institute of Particle Physics, Central China Normal University, Wuhan 430079, China}
\author{Wei Li}
\affiliation{Key Laboratory of Quark and Lepton Physics (MOE) and Institute of Particle Physics, Central China Normal University, Wuhan 430079, China}
\affiliation{Author to whom correspondence should be addressed: liw@mail.ccnu.edu.cn}
\author{Yuxiang Yang}
\affiliation{Key Laboratory of Quark and Lepton Physics (MOE) and Institute of Particle Physics, Central China Normal University, Wuhan 430079, China}

\begin{abstract}
The rich-club phenomenon, which provides information about the association between nodes, is a useful method to study the hierarchy structure of networks. In this work, we explore the behavior of rich-club coefficient (RCC) in scale-free networks, and find that the degree-based RCC is a power function of degree centrality with power exponent $\beta$ and the betweenness-based RCC is a linear function of betweenness centrality with slope $\theta$. Moreover, we calculate the value of RCC in a BA network by deleting nodes and obtain a general expression for RCC as a function of node sequence. On this basis, the solution of RCC for centrality is also obtained, which shows that the curve of RCC is determined by the centrality distribution. In the numerical simulation, we observe: $\beta$ and $\gamma$ (the degree distribution exponent) increase together, $\theta$ increases with the average degree and decreases to convergence as $\gamma$ increases.\par

\end{abstract}

\maketitle

\section{Introduction}
\label{intro}

\tcb{Scale-free networks in the real world, such as social networks, transportation networks, and biological networks, often exhibit heterogeneous connectivity and structured organization, resulting in the emergence of modules, hierarchies, high-order structures and so on \cite{fortunato2022,de2023more,battiston2021physics,boguna2021network}, affecting their resilience and functionality across various domains \cite{artime2024robustness}. The way nodes are connected also introduces distinctions among network nodes.} In a large network, the influence of each node is different, with some nodes holding more significant influence and acting as “hubs”. These hubs control information flow and have great impacts on transportation efficiency of the network, which means that the hubs are critical to the overall structure and function of the network \cite{opsahl2008prominence}. Therefore, \tcb{the robustness and resilience of networks show significant responses to these hubs,} and deliberate attacks on hubs can result in the collapse of the entire network \cite{artime2024robustness}. If there are more dense connections within hubs than between hubs and other nodes, that is, hubs are more likely to connect with one another, then a rich-club will be formed \cite{zhou2004rich}. \tcb{The rich-club is also a type of mesoscale structure, and more accurately described as a core structure \cite{gallagher2021clarified}.} This phenomenon can be observed across many types of networks, such as social networks \cite{ansell2016says,dong2015inferring,vaquero2013rich}, transport networks \cite{wei2018rich,li2007empirical,zhang2021unveiling,zhu2021exploring}, and biological networks \cite{van2011rich,ball2014rich}, etc. Essentially, the rich-club phenomenon provides insight into the strong connectivity between rich nodes (hubs). The close connection between rich nodes is the key factor behind their significant influence \cite{jiang2008statistical}. This facilitates the rapid dissemination of influence among them, ultimately affecting the entire network \cite{berahmand2018effect}. A strong rich-club in a social network suggests that the network is not composed of disjointed and loosely connected small groups. Rather, there is a dominant “top-level” group that serves as the core structure of the network \cite{cinelli2017resilience}. Intuitively, nodes with higher degree have more edges, and connect to each other more easily, thus forming a more densely connected club. \par

The rich-club coefficient (RCC) is a network metric that measures the normalized density of connections among rich nodes in a network. It quantifies the tendency for these nodes to form a densely interconnected rich club. RCC, denoted by $\phi(r)$, is the ratio of the actual number of edges between nodes to the maximum possible number of edges that could exist between them \cite{zhou2004rich}. $\phi(r)$ is given by,
\begin{equation}
\begin{split}
\phi(r)=\dfrac{2E_{>r}}{N_{>r}(N_{>r}-1)}\;,
\end{split}
\label{eqs:1}
\end{equation}
where $r$ is the richness that describes the importance of nodes, such as degree or betweenness centrality. $N_{>r}$ is the number of nodes whose richness is greater than $r$, and $E_{>r}$ is the total number of edges among these nodes. The coefficient $\phi(r)$ measures connectivity and ranges from 0 to 1, with higher values indicating greater connectivity. For example, $\phi(r)=1$ represents that the top ranked r nodes are fully connected. However, the detection of rich-club phenomenon in a network cannot only rely on the growth process of RCC. It has been demonstrated that a randomly produced Erd\H{o}s–R\'{e}nyi (ER) random network also displays a consistently increasing RCC \cite{colizza2006detecting}. This finding emphasizes that an increase in the coefficient value with node centrality does not always indicate the presence of the rich-club phenomenon in the network. Therefore the normalized rich-club coefficient is proposed, which is defined as $\rho_{ran}(r)=\phi(r)\slash\phi_{ran}(r)$. $\phi_{ran}(r)$ is RCC of 1-order null model. When the coefficient $\rho_{ran}(r)$ is greater than 1 in the high richness region, the rich club phenomenon is obvious in the network.\par

\begin{figure*}[htbp]
	\begin{tabular}{cc}   
		\includegraphics[scale=0.32]{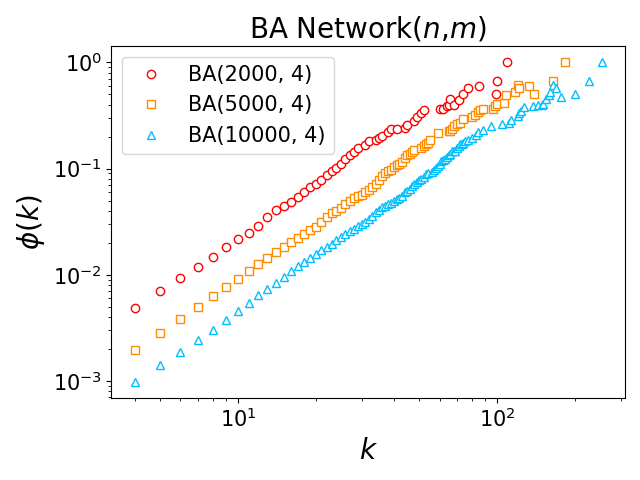} &
		\includegraphics[scale=0.32]{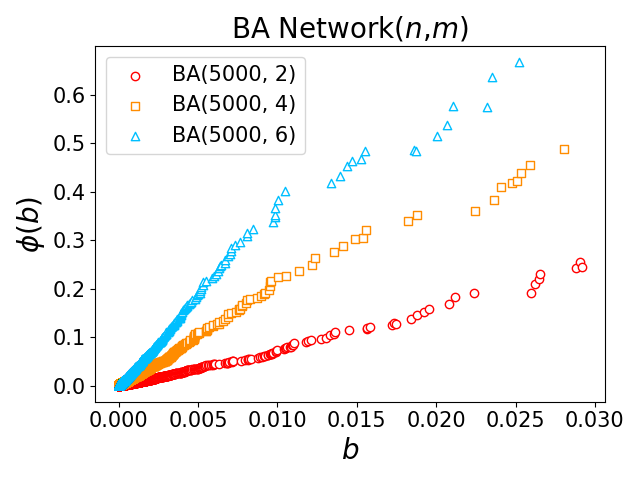} \\
		\text{(a)} & \text{(b)}
	\end{tabular}
	\caption{  The value of RCC in BA network BA$(n,m)$, where $n$ is the number of nodes and $m$ is the number of new edges introduced at each step. \tcb{Due to numerous and contiguous nodes of the same type, we could intuitively observe the linear relation in both graphs.}  (a) Log-log plot of $\phi(k)$ versus degree centrality. (b) The plot of $\phi(b)$ versus betweenness centrality. As the maximum value of betweenness centrality is too large, for displaying the feature, the top one percent of nodes has been removed from the plot.  }
	\label{Fig.1}
\end{figure*}

\begin{figure*}[htbp]
	\begin{tabular}{cc}   
		\includegraphics[scale=0.32]{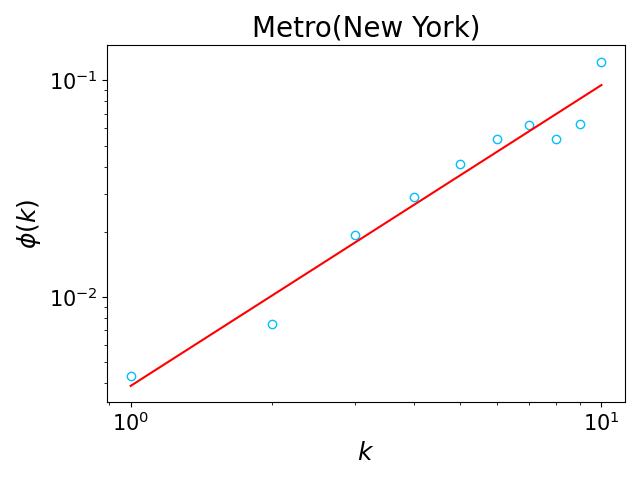} &
		\includegraphics[scale=0.32]{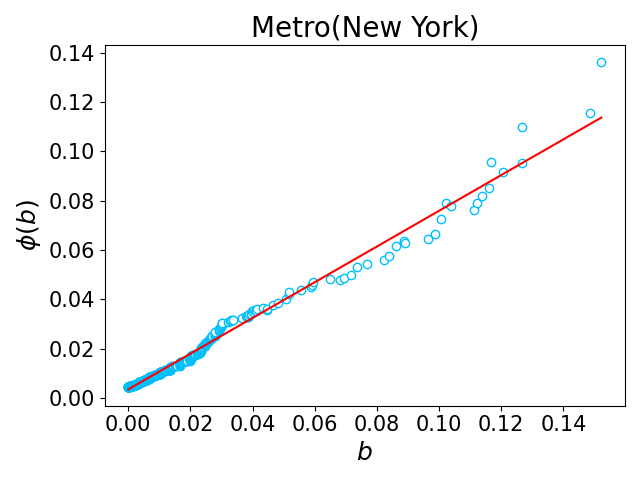} \\
		\text{(a)} & \text{(b)}
	\end{tabular}
	\caption{   The value of RCC in New York metro network, and the conclusions drawn from the graph also satisfy Eq. (\ref{eqs:2}) well. (a) Log-log plot of $\phi(k)$ versus degree centrality. Even with a small number of degree values, a good fit to a straight line can still be achieved. (b) The plot of $\phi(b)$ versus betweenness centrality. }
	\label{Fig.2}
\end{figure*}

The “fat tail” of the power-law degree distribution in scale-free networks indicates that some high-degree nodes have many edges \cite{hein2006scale}. This makes it possible to have more connections between high-degree nodes. Consequently, the occurrence of the rich-club phenomenon can be observed in scale-free networks \cite{colizza2006detecting}. In our study, we find a power-law relation between RCC and degree centrality. Moreover, when richness is set to be betweenness centrality, RCC is linearly related to betweenness centrality. The model we use to show the results is the Barab{\'a}si-Albert model \cite{barabasi1999emergence}, the so called BA network, which is a representative model for scale-free networks. Here the node-betweenness centrality is rescaled into the unit interval (0,1) in our calculations. As shown in Fig.\ref{Fig.1}, the degree-based RCC $\phi(k)$ versus $k$ can be fitted by a straight line in log-log coordinates, and the betweenness-based RCC $\phi(b)$ versus $b$ can be fitted by a straight line in linear coordinates. Therefore, the two relations can be written as, 
\begin{equation}
\begin{split}
\ \phi(k) \sim k^{\beta}   \;,
\end{split}
\begin{split}
\ \phi(b) \sim \theta \cdot b \;.
\end{split}
\label{eqs:2}
\end{equation}
Where $\beta$ is the power exponent of $\phi(k)$, and $\theta$ is the slope of $\phi(b)$. The phenomenon can also be observed in real networks, as evidenced by the New York metro network, which is the largest subway network in the world and exhibits scale-free characteristics \cite{derrible2010complexity}. Similar results can be obtained from Fig.\ref{Fig.1} and Fig.\ref{Fig.2}, and it is worthy to note that the degree value of the metro network is relatively small due to its structural constraints, resulting in sparser nodes in Fig.\ref{Fig.2}(a).

The paper is organized as follows. Sec. II introduces the functional relation between degree and betweenness in scale-free networks \tcb{by referencing relevant literature}. In Sec. III, we employ two methods to calculate RCC, \tcb{and explore what factors contribute to the relation between RCC and node centrality.}\tcb{In Sec. IV, we observe that $\beta$ and $\theta$ exhibit good behaviors in numerical simulations of scale-free networks, and we present and discuss these findings accordingly.} Finally, we conclude in Sec. V.

\section{Centrality in scale-free networks}
The most prominent characteristic of a scale-free (SF) network is its degree distribution, which follows the power law, $p(k) \sim k^{-\gamma}$. This feature is widely recognized as key to identify SF networks \cite{cohen2010complex}. For BA network, it is a growth model with preferential linking, and yields a power-law degree distribution with $\gamma=3$ \cite{barabasi1999emergence} . Degree centrality (DC) describes the importance of a node based on the number of edges connected to it. In contrast, betweenness centrality (BC) represents the impact of a node on communication between every pair of vertices. Nodes with larger BC will be on more shortest paths. This makes BC a useful tool for network topology analysis. Goh \textit{et al}. have shown that the BC (or “load”) distribution in a SF network obeys power-law with exponent $\delta$ \cite{goh2001universal}: $p(b) \sim b^{-\delta}$ (Since the values of BC are continuous, we here only obtain the cumulative distribution of $P(b) \sim b^{1-\delta}$). There is a strong connection between DC and BC. Especially in SF networks, nodes with high degree values typically create an obvious core structure, which is in the center of communication between nodes \cite{ma2015rich}. Due to the discreteness of degree values, nodes with fixed degree value will have more than one betweenness value. To avoid this, we set $b_{k}$ as the average BC value of each group of nodes with degree k, and can obtain the power-law relation between $b(k)$ and $k$ \cite{goh2001universal}: 
\begin{equation}
\begin{split}
\ b_{k} \sim k^{\eta}\;.
\end{split}
\label{eqs:3}
\end{equation}
Such a relation describes the correlation between DC and BC in scale-free networks. Transforming it into the cumulative distribution \cite{vazquez2002large}, one obtains,
\begin{equation}
\begin{split}
\ P_{k}(k) \sim k^{\eta(1-\delta)} \sim k^{(1-\gamma)}\;,
\end{split}
\label{eqs:4}
\end{equation}
which confirms the following scaling relation
\begin{equation}
\begin{split}
\ \eta = \dfrac{\gamma-1}{\delta-1}\;.
\end{split}
\label{eqs:5}
\end{equation}
According to Barth{\'e}lemy \cite{barthelemy2004betweenness}, the values of $\delta$ and $\eta$ are not universal and depend on specific details of the networks. However, for $m=1$ in a BA network, it has been shown in a tree network that $\eta = \delta =2$ \cite{barthelemy2004betweenness,masoomy2023relation}.\par

In the previous section we have known that RCC is a power function of DC and a linear function of BC, given by $\phi(k) \sim k^{\beta}$ and $\phi(b) \sim \theta \cdot b$, respectively. Comparing them to the relation (3) leads to,
\begin{equation}
\begin{split}
\ \eta \approx \beta\;.
\end{split}
\label{eqs:6}
\end{equation}
It can be checked, as also shown in Fig.\ref{Fig.2}, that the slopes of two fitted lines are nearly identical in a log-log plot. In SF networks, understanding DC and BC is crucial for understanding the characteristics of networks. From the distribution of centrality to the association between centrality, scholars have been exploring various new features to describe networks more accurately. The emergence of rich-club theory undoubtedly provides a new method to study the association between nodes based on centrality, and brings answers to the macro phenomena of the network.

\section{Analysis of rich-club coefficient}
\subsection{RCC and Correlation}

In order to verify the curve of RCC in SF networks and explore how the relation between RCC and centrality is generated. We begin by dismantling the network similarly to the k-core method \cite{dorogovtsev2006k}, and estimate the number of edges between nodes at each level. When the richness is DC, the form of RCC is to summarize the number of edges between nodes of different degrees. The number of edges in each group provides the data required for calculating the joint degree distribution, which establishes a relation between RCC and joint distribution. Therefore, we can use this to calculate RCC. The joint degree distribution of any given network is
\begin{equation}
\begin{split}
\ p(i,j) = \dfrac{E_{i,j}}{N\left \langle k \right \rangle }\;,
\end{split}
\label{eqs:7}
\end{equation}
where $E_{i,j}$ is the number of edges between nodes with degree $i$ and nodes with degree $j$, $N$ is the number of nodes, and $\left \langle k \right \rangle $ is the average degree of the network. Then we can obtain, $2E_{>k} = \textstyle \sum_{ij}E_{ij} = \textstyle \sum_{ij}N\left \langle k \right \rangle p(i,j), N_{>k} = N\textstyle \sum_{i}p(i)$, \tcb{where $p(i)$ is the degree distribution}. Converting the numerator and denominator of $\phi(k)$ into calculating cumulants, we obtain,
\begin{equation}
\begin{split}
\ \phi(k) = \frac{ {\textstyle \sum_{ij}N\left \langle k \right \rangle p(i,j)} }{N {\textstyle \sum_{i}p(i)}[N {\textstyle \sum_{i}p(i)}-1] } \;.
\end{split}
\label{eqs:8}
\end{equation}
It shows clearly that $\phi(k)$ is a cumulants of the joint degree distribution, and the connection between nodes calculated by $\phi(k)$ can be obtained by the sum of degree-degree correlations. Therefore, RCC is also a method to calculate correlations, with a larger RCC indicating a strong correlation.\par

The calculation of joint degree distribution is commonly divided into two cases. One is that in uncorrelated networks, the joint degree distribution can be decomposed as
\begin{equation}
\begin{split}
\ p_{unc}(i,j)=\frac{ijp(i)p(j)}{\left \langle k \right \rangle ^2}  \;.
\end{split}
\label{eqs:9}
\end{equation}
In this case, the joint degree distribution is a function of the degree distribution. If $k$ is assumed to be a continuous variable \cite{colizza2006detecting}, in the approximation of $k_{max} \to \infty $, L'H{\^o}pital's rule can give $\phi(k)$ in a simplified version,
\begin{equation}
\begin{split}
\              
\phi (k)_{unc} \approx  \frac{\iint_{k}^{k_{max}} ijp(i)p(j)didj }{N\left \langle k \right \rangle [\int_{k}^{k_{max}}p(i)di ]^2 }\sim \frac{k^2}{N\left \langle k \right \rangle}   
\;.
\end{split}
\label{eqs:10}
\end{equation}
The result shows that the power function also applies to $\phi(k)$ in uncorrelated networks under the approximation of $k_{max} \to \infty $.\par

For SF networks, the degree-degree correlation has been studied \cite{newman2002assortative}. In particular, degree-degree correlation also exists in the BA network if the assortativity coefficient $R$ is close to $0$. Due to the preferential attachment, nodes are more likely to connect to high-degree nodes, which means that the connections between nodes will be affected by degree values. It has been proven in \cite{krapivsky2001organization} that the joint degree distribution of BA($m=1$) networks does not conform to Eq. (\ref{eqs:9}). And the simulation results \cite{li2011emergence,noldus2015assortativity} have shown that $R$ is changeable in the BA network. That the assortativity coefficient $R$ approaches $0$ in the BA network does not necessarily mean that nodes are connected randomly. Compared to high-degree nodes, connections between low-degree nodes are much sparser. In conclusion, the joint degree distribution of the SF networks remains a challenge. The analytical expressions obtained that have been obtained are very complicated \cite{fotouhi2013degree,pekoz2017joint}, which will complicate our problems. This makes us give up the idea of using the joint distribution to calculate RCC of SF networks. But it does not mean we neglect the contribution of the joint degree distribution, which plays a valuable role in helping us understand that RCC represents an accumulation of correlation.

\begin{figure}
\centering
\includegraphics[width=0.4\textwidth]{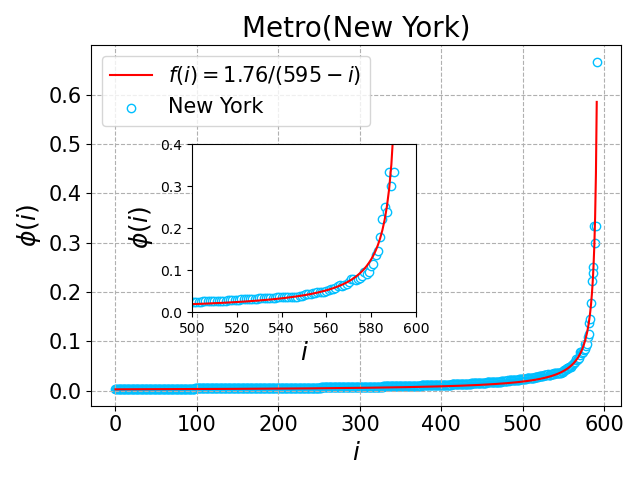}
\caption{ The inverse proportional relation between RCC $\phi (i)$ and betweenness-based node sequence $i$. \tcb{We successfully validated the inverse proportion in Eq. (\ref{eqs:13}) using the New York metro network.} The value of $\phi (i)$ overlaps well with the curve $f(i)= \frac{1.76}{595-i}$, and we also show local details in the subplot.}
\label{Fig.3}
\end{figure}

\begin{figure}
\centering
\includegraphics[width=0.4\textwidth]{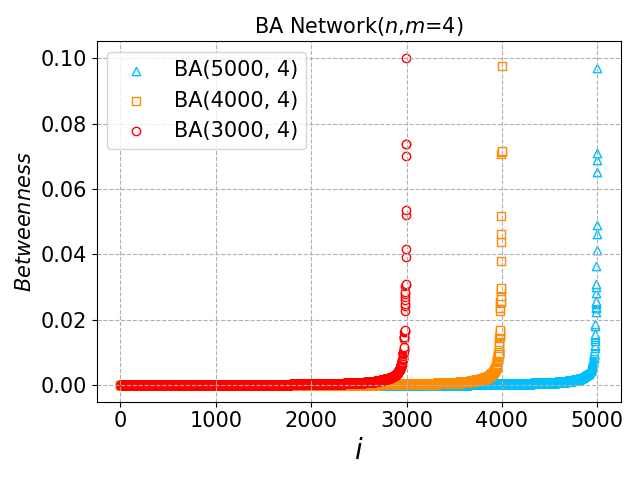}
\caption{ Sequences of betweenness centrality in BA networks. The  nodes are sorted by their values of BC. The abscissa is the sequence number, and the ordinate is the corresponding value of BC. \tcb{We can find that the curve of BC and sequence number $i$ also follows an inverse proportional function, expressed as $b_{i}\sim \frac{1}{n-i}$.}}
\label{Fig.4}
\end{figure}

\begin{figure*}[htbp]
	\begin{tabular}{cccc}   
		\includegraphics[scale=0.195]{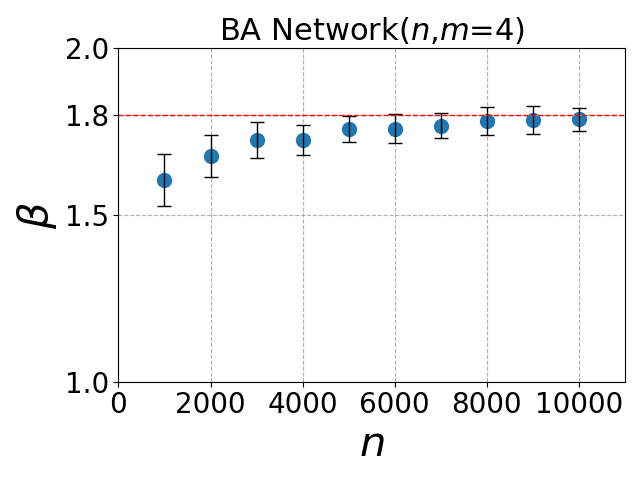} &
		\includegraphics[scale=0.195]{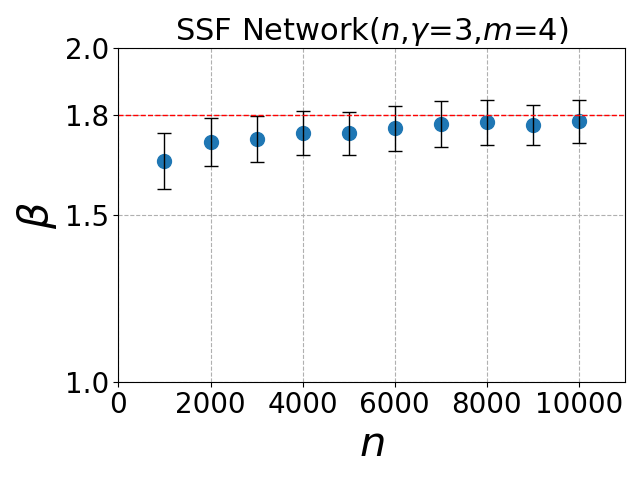} &
		\includegraphics[scale=0.195]{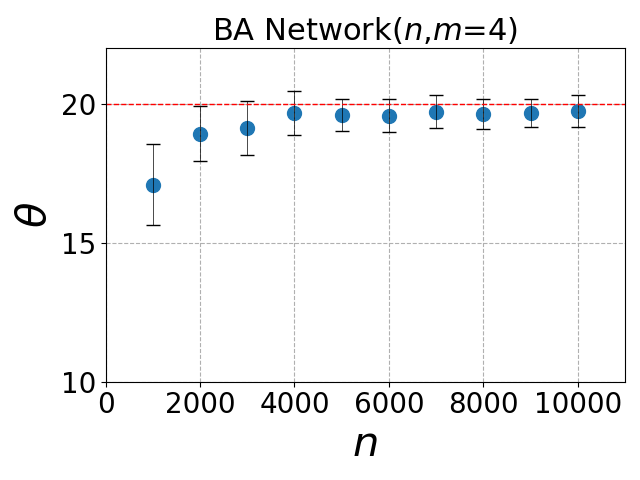} &
		\includegraphics[scale=0.195]{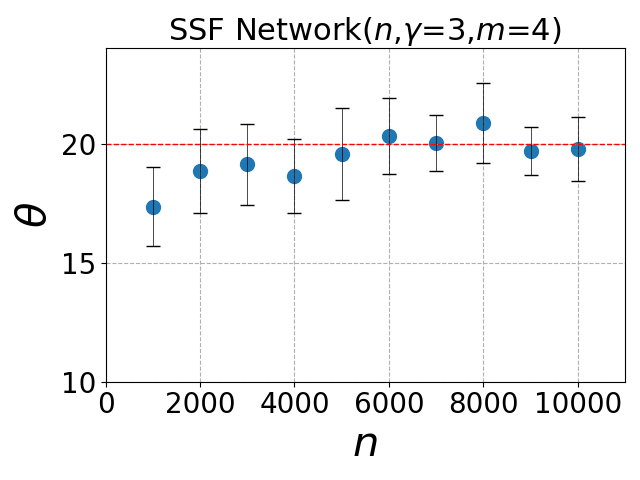} \\
		\text{(a)} & \text{(b)} & \text{(c)} & \text{(d)}
	\end{tabular}
	\caption{  Values of $\beta$ and $\theta$ in scale-free networks under different $n$, but with the same $\gamma=3$ and $m=4$. (a)(b) Power exponent $\beta$ for degree-based RCC. (c)(d) Slope $\theta$ for betweenness-based RCC. Both $\beta$ and $\theta$ converge to a value for large $n$.}
	\label{Fig.5}
\end{figure*}

\subsection{RCC and Centrality Distribution}
\tcb{To elucidate how the relation between RCC and centrality is determined, we proceed to investigate the RCC of a SF network through an evolutionary process.}  Let's consider a BA network BA$(n,m)$ with $m$ new edges being added at each step. \tcb{The network has $n$ nodes and $E$ edges.} Nodes in the network are ranked by their richness in ascending order, and each node is assigned a sequence number based on its position in the ranking list. \tcb{Here, $i$ and $j$ are used to denote the sequence numbers of nodes. As we remove nodes in order until node $i$ is removed,} the network density is
\begin{equation}
\begin{split}
\ \rho_{>i} = \dfrac{2(E- {\textstyle \sum_{1}^{i}l_{j}} )}{(n-i)(n-i-1)}\;.
\end{split}
\label{eqs:11}
\end{equation}
The variable $l_j$ represents the number of edges that are removed when node \tcb{$j$} is cut off from the network. The density calculated using this formula serves as RCC of the network, but its calculation poses a challenge due to the varying values of $l_j$ with node deletion at each step. Thus, computing the accurate RCC value is a difficult task. When the richness is DC, preferential linking associates the sequence in which nodes are removed with the same sequence in which nodes join the network. As the mean of $l_{j}$ in the BA network is $m$, using $l_{j}=m$ directly will not affect our conclusion. In this case, the total number of edges $E= {\textstyle \sum_{1}^{n-m}l_{j}} +m$. And RCC is given by
\begin{equation}
\begin{split}
\  \phi (i)=\frac{ 2({\textstyle \sum_{i+1}^{n-m}}l_{j}+m) }{(n-i)(n-i-1)}=\frac{2(n-m-i)m}{(n-i)(n-i-1)}  \;.
\end{split}
\label{eqs:12}
\end{equation}
When $i<n-m, n\gg m$
\begin{equation}
\begin{split}
\  \phi (i) \approx \frac{2m(n-i)}{(n-i)(n-i)} \approx \frac{2m}{n-i}  \;,
\end{split}
\label{eqs:13}
\end{equation}
which is an inverse proportional function. The function holds for any sequence of centrality that is positively related to DC. This relation between $\phi (i)$ and betweenness-based $i$ can also be observed in the New York metro network, as shown in Fig.\ref{Fig.3}, where the real $\phi (i)$ fits the fitted curve $f(i)= \frac{1.76}{595-i}$ very well. In addition, we enlarge the details of the circle points and fitted curves in the subgraph.\par

\begin{figure*}[htbp]
	\begin{tabular}{cccc}   
		\includegraphics[scale=0.195]{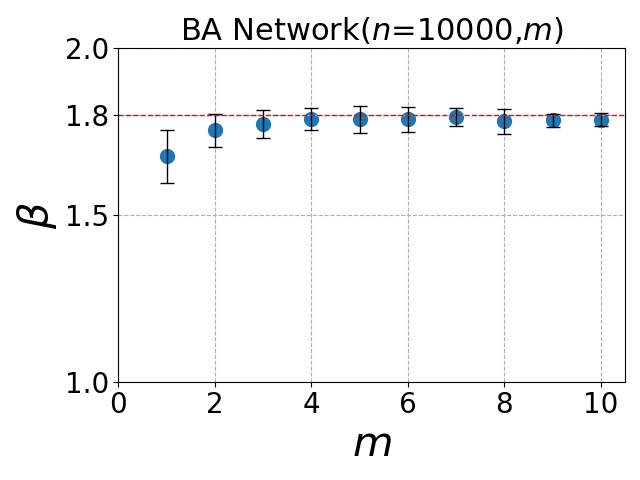} &
		\includegraphics[scale=0.195]{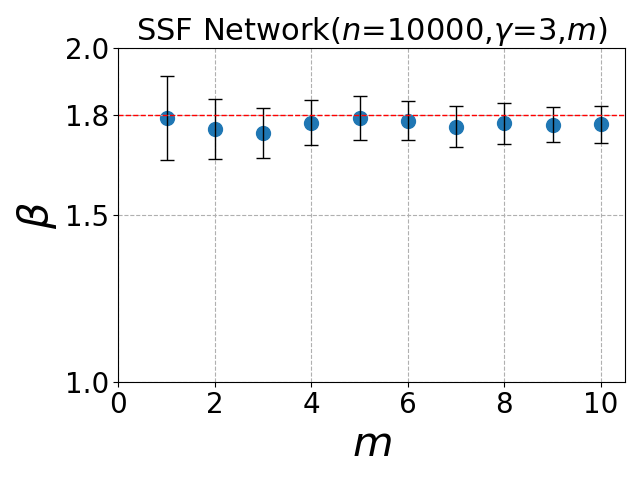} &
		\includegraphics[scale=0.195]{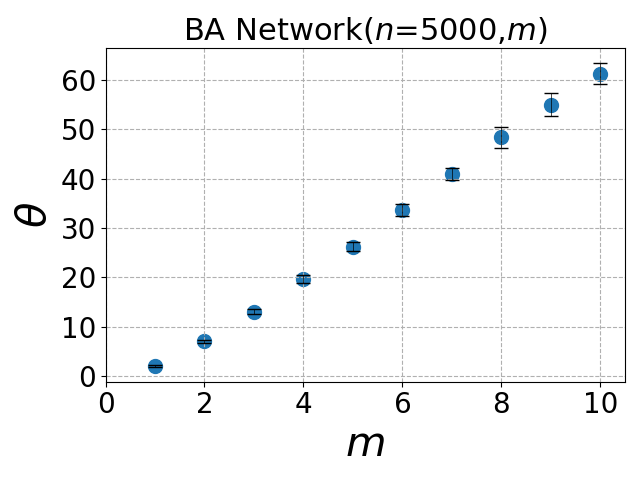} &
		\includegraphics[scale=0.195]{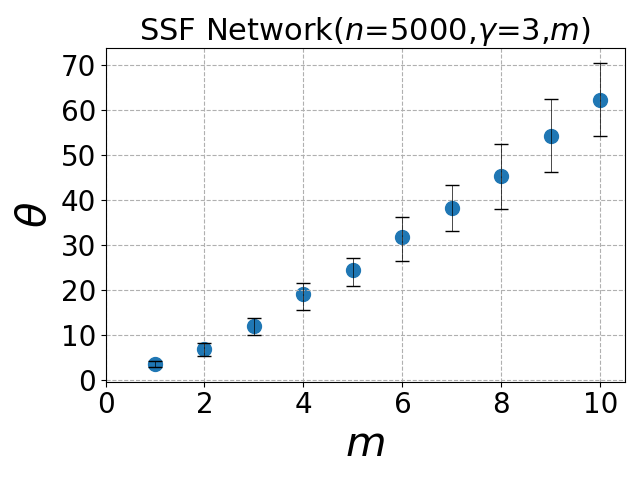} \\
		\text{(a)} & \text{(b)} & \text{(c)} & \text{(d)}
	\end{tabular}
	\caption{ Values of $\beta$ and $\theta$ in scale-free networks under different $m$ values, but with the same $n$ and $\gamma=3$. The figures (a) and (b) indicate that the value of $\beta$ is near $1.8$. The figures (c) and (d) show that $\theta$ and $m$ have linear correlations. }
	\label{Fig.6}
\end{figure*}

\begin{figure*}[htbp]
	\begin{tabular}{cc}   
		\includegraphics[scale=0.3]{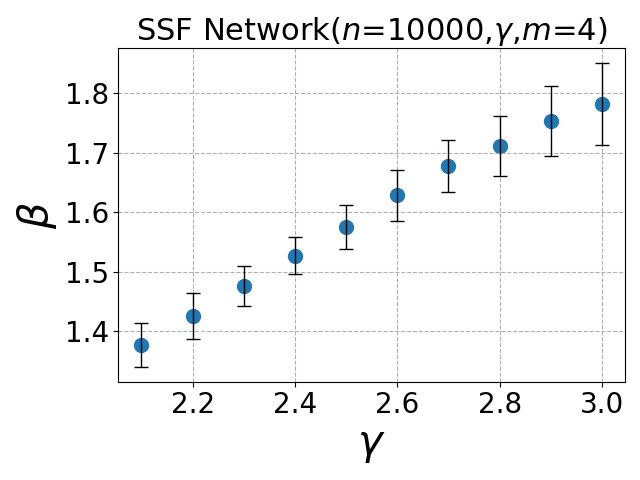} &
		\includegraphics[scale=0.3]{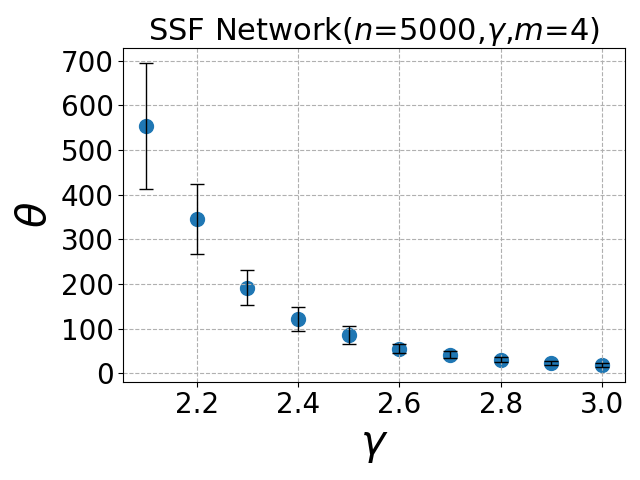} \\
		\text{(a)} & \text{(b)}
	\end{tabular}
	\caption{ Exponent $\beta$ and slope $\theta$ versus $\gamma$ in SSF networks. (a) Linear relation between $\beta$ and $\gamma$. (b) Inverse relation between $\theta$ and $\gamma$. }
	\label{Fig.7}
\end{figure*}

Next, we aim to discuss the relation between centrality and node sequence, and attempt to obtain RCC of centrality. When the richness is BC, due to the proportional relationship between BC and DC in Eq. (\ref{eqs:3}), the sequence of BC can also satisfy the relation given by Eq. (\ref{eqs:13}). And, the curve of BC and sequence number $i$ shown in Fig.\ref{Fig.4} also follows an inverse proportional function, expressed as $b_{i}\sim \frac{1}{n-i}$. Combining the formula in Eq. (\ref{eqs:13}), a linear relation can be obtained: $\phi (b) \sim \theta \cdot b$. It shows that the linear relation of RCC is the result of the contrast between two inverse proportional functions, with the distribution of BC being the main factor. When the richness is DC, as $\phi(k)$ is the average value for a given degree, we are unable to reveal the relation between $\phi(k)$ and $k$ through the relation between $k$ and $i$. Instead, calculating $\phi(k)$ requires using $b_{k} \sim k^{\eta}$, which also takes an average result, and ultimately leads to a new relation in the form of $\phi (k)\sim k^\beta$. And relation (3) combines two centrality distributions \cite{goh2001universal}, so $\phi(k)$ is also determined by them. In this section, we have further verified that the power function between RCC and DC, as well as the linear relation between RCC and DC, exist in SF networks. Additionally, we get the conclusion that the distribution of centrality determines the characteristics of the RCC curve. According to the calculation process above, the degree distribution may be the main factor. Therefore, the relation (2) is universal in SF networks. And we can also predict that the values of $\beta$ and $\theta$ in the SF network will be affected by the exponent $\gamma $.

\section{The behaviors of exponent $\beta$ and slope $\theta$}
After understanding how the curve of RCC is formed, our goal is to explore what affects the values of $\beta$ and $\theta$. We will study the parameters of network construction, which are the most basic data of a network. These include the number of nodes ($n$), the number of new edges ($m$), and the degree distribution exponent($\gamma$) in SF networks. In the numerical calculation, we use two SF models for comparison and verification. One is the BA network, generated through dynamic growth with adjustable parameters $n$ and $m$. For comparison, we construct a static scale-free (SSF) network based on the degree sequence $\left \{ k_{n} \right \} $ following a power-law distribution \cite{platt2019network,bayati2010sequential}. Unlike the BA network, the SSF network does not have a dynamic growth process, and has adjustable $\gamma$ (We set $\gamma$ in the usual range $\left ( 2,3 \right ] $). And we can multiply the degree sequence $\left \{ k_{n} \right \} $ by $m$ to change the network density. It also increases the average degree by a factor of $m$. Therefore, the SSF network is adjustable with parameters $n$, $m$ and $\gamma$.

Initially, we discuss the effect of the change in the number of nodes on the exponent $\beta$ and slope $\theta$. To enable the smoothness of fitting, we base our results on the average outcomes of multiple calculations. As shown in Fig.\ref{Fig.5}, we construct the BA network and SSF network with the same $m=4$ and $\gamma=3$, and then gradually increase the number of nodes. We can see the values of $\beta$ and $\theta$ rise to convergence as $n$ increases, and conclude that the values of exponent $\beta$ and slope $\theta$ are not affected by the number of nodes in a large SF network. This also verifies that the exponent $\eta$ does not vary with network size \cite{barthelemy2004betweenness}. In addition, when $n$ is small, both $\beta$ and $\theta$ rise to convergence as an inverse proportional function. This is because increasing $n$ does not change the degree distribution but decreases the initial density $\phi(0)$ ($\phi(0) \sim \frac{1}{n}$). So the fitted $\beta$ changes with $\phi(0)$ together, and increases until it converges. Similar to the case of DC, when richness is BC, the slope $\theta$ has the same converging behavior.

In Fig.\ref{Fig.6}, we have kept $n$ and $\gamma$ constant while setting $m\in \left [ 1,10 \right ]$. Our results indicate distinct behaviors between $\beta$ and $\theta$. Specifically, we have found that the exponent $\beta$ is relatively stable across the range of $m$, whereas the slope $\theta$ exhibits a linearly positive correlation with $m$. As $m$ directly affects the density of the network, it is inconsistent with our experience that the fitting value $\beta$ remains constant with changes in $m$. This may be the fact that $m$ is too small for $n$ in the usual range, resulting in little change in initial density. Unlike the case in exponent $\beta$, the value of $\theta$ is impacted by $\phi (i) \approx \frac{2m}{n-i}$, resulting in a proportional increase between $\theta$ and $m$. However, due to the limitation of fitting accuracy, we have only explained the trend of the curve without discussing the obtained numerical result in detail. But our numerical results also make sense, they show $\beta \approx 1.8 , \delta \approx 2.1$, which is in the proximity of the previous one \cite{goh2001universal,masoomy2023relation}.  \par


Finally, we change the degree distribution exponent $\gamma$ in SSF network. Fig.\ref{Fig.7} shows when $\gamma$ increases in the interval $\left ( 2,3 \right ] $, $\beta$ and $\theta$ also exhibit different behaviors. $\beta$ and $\gamma$ increase together, they tend to increase linearly; $\theta$ gradually decreases to convergence as $\gamma$ increases. The exponent $\gamma$ changes the value of the degree sequence $\left \{ k_{n} \right \} $ and affects the distribution of BC in SSF networks. However, the precise manner in which $\gamma$ affects the details of the network remains unclear. We cannot find the reason that $\beta$ and $\gamma$ are linearly dependent, and why $\theta$ converges. It is a very worthy problem to discuss, which may require a deeper understanding of node centrality.

\section{Summary and Conclusions}
In the previous section, we have studied the behavior of RCC in scale-free networks, and found that RCC is an important quantity to describe the characteristics of scale-free networks. It reveals the connection between rich nodes through the centrality, and own a association process with the distribution of centrality. The numerical results show that the relation between RCC and DC is a power function $\phi(k) \sim k^{\beta}$, and the relation between RCC and BC is a linear function $\phi(b) \sim \theta \cdot b$. The results from the New York metro network, which exhibits scale-free properties, also confirm our findings. Next, compared to the existing centrality features in scale-free networks, we also find that the exponents $\beta$ and $\eta $ are close in value. Furthermore, RCC offers a method for calculating correlation through the representation of cumulative joint distribution. For example, when richness is degree, a larger RCC means not only greater density but also greater degree-degree correlation among a group of nodes. In order to explore how the RCC is formed, we first try to calculate $\phi (k)$ with the aid of the joint degree distribution. But we did not succeed, because the current results of the joint degree distribution do not simplify our calculations, it cannot help us find the factors affecting the curve. Then we changed the method and tried to calculate RCC in scale-free networks by removing nodes. In BA networks, we calculate the result of RCC and node sequence, and under the approximation of removing $m$ edges in every step, the simplified relation is $\phi (i) \approx \frac{2m}{n-i}$. This is a general result, where $i$ can belong to any sequence of centrality (or other data \cite{cinelli2019generalized}) positively related to DC. The sequence of BC is used, and it shows that the sorting of BC is an inverse proportional function. So the linear function of $\phi (b)$ is the result of two inverse proportional functions. Because $\phi (k)$ is the average result under the same degree value, we use relation (3) to calculate and get the power function. It can be concluded that the relation between RCC and centrality is dominated by the distribution of centrality, and the sequence of centrality significantly affects the curve of RCC. As such, understanding the characteristics of RCC is an import component of comprehending scale-free networks.\par

We have also discussed what affect the values of $\beta$ and $\theta$ in terms of network construction parameters. In order to compare the results of BA networks, we use a static scale-free network with strong randomness, whose degree distribution is randomly generated for a given exponent $\gamma$. In many simulations, the results show that $\beta$ and $\gamma$ increase together, $\theta$ increases with $m$ (related to the average degree) and decreases to convergence as $\gamma$ increases. To understand why $\beta$ and $\theta$ behave this way, we need to have a deeper understanding of centrality and explore how the change of $\gamma$ changes the network structure. Moreover, the study of joint distribution in scale-free networks can also help us further understand the behavior of RCC.\par

\textbf{Acknowledgements} 
This work was supported in part by the Fundamental Research Funds for the Central Universities, China (Grant No. CCNU19QN029), and the National Natural Science Foundation of China (Grant No. 61873104).

\tcb{\textbf{Code availability} 
The code used for the performance comparisons can be found in the 
repository https://github.com/angryRPG/RCC-in-SF-network.}

\subsection*{References}

\bibliographystyle{unsrtnat}
\bibliography{rich-club}

\providecommand{\noopsort}[1]{}\providecommand{\singleletter}[1]{#1}%
\begin{thebibliography}{40}
\providecommand{\natexlab}[1]{#1}
\providecommand{\url}[1]{\texttt{#1}}
\expandafter\ifx\csname urlstyle\endcsname\relax
  \providecommand{\doi}[1]{doi: #1}\else
  \providecommand{\doi}{doi: \begingroup \urlstyle{rm}\Url}\fi

\bibitem[Fortunato and Newman(2022)]{fortunato2022}
Santo Fortunato and Mark~EJ Newman.
\newblock 20 years of network community detection.
\newblock \emph{Nature Physics}, 18\penalty0 (8):\penalty0 848--850, 2022.

\bibitem[De~Domenico(2023)]{de2023more}
Manlio De~Domenico.
\newblock More is different in real-world multilayer networks.
\newblock \emph{Nature Physics}, 19\penalty0 (9):\penalty0 1247--1262, 2023.

\bibitem[Battiston et~al.(2021)Battiston, Amico, et~al.]{battiston2021physics}
Federico Battiston, Enrico Amico, et~al.
\newblock The physics of higher-order interactions in complex systems.
\newblock \emph{Nature Physics}, 17\penalty0 (10):\penalty0 1093--1098, 2021.

\bibitem[Boguna et~al.(2021)Boguna, Bonamassa, et~al.]{boguna2021network}
Marian Boguna, Ivan Bonamassa, et~al.
\newblock Network geometry.
\newblock \emph{Nature Reviews Physics}, 3\penalty0 (2):\penalty0 114--135, 2021.

\bibitem[Artime et~al.(2024)Artime, Grassia, et~al.]{artime2024robustness}
Oriol Artime, Marco Grassia, et~al.
\newblock Robustness and resilience of complex networks.
\newblock \emph{Nature Reviews Physics}, 6\penalty0 (2):\penalty0 114--131, 2024.

\bibitem[Opsahl et~al.(2008)Opsahl, Colizza, et~al.]{opsahl2008prominence}
Tore Opsahl, Vittoria Colizza, et~al.
\newblock Prominence and control: the weighted rich-club effect.
\newblock \emph{Physical review letters}, 101\penalty0 (16):\penalty0 168702, 2008.

\bibitem[Zhou and Mondrag{\'o}n(2004)]{zhou2004rich}
Shi Zhou and Ra{\'u}l~J Mondrag{\'o}n.
\newblock The rich-club phenomenon in the internet topology.
\newblock \emph{IEEE communications letters}, 8\penalty0 (3):\penalty0 180--182, 2004.

\bibitem[Gallagher et~al.(2021)Gallagher, Young, and Welles]{gallagher2021clarified}
Ryan~J Gallagher, Jean-Gabriel Young, and Brooke~Foucault Welles.
\newblock A clarified typology of core-periphery structure in networks.
\newblock \emph{Science advances}, 7\penalty0 (12):\penalty0 eabc9800, 2021.

\bibitem[Ansell et~al.(2016)Ansell, Bichir, and Zhou]{ansell2016says}
Christopher Ansell, Renata Bichir, and Shi Zhou.
\newblock Who says networks, says oligarchy? oligarchies as “rich club” networks.
\newblock \emph{Connections}, 36\penalty0 (1):\penalty0 20--32, 2016.

\bibitem[Dong et~al.(2015)Dong, Tang, et~al.]{dong2015inferring}
Yuxiao Dong, Jie Tang, et~al.
\newblock Inferring social status and rich club effects in enterprise communication networks.
\newblock \emph{PloS one}, 10\penalty0 (3):\penalty0 e0119446, 2015.

\bibitem[Vaquero and Cebrian(2013)]{vaquero2013rich}
Luis~M Vaquero and Manuel Cebrian.
\newblock The rich club phenomenon in the classroom.
\newblock \emph{Scientific reports}, 3\penalty0 (1):\penalty0 1--8, 2013.

\bibitem[Wei et~al.(2018)Wei, Song, et~al.]{wei2018rich}
Ye~Wei, Wei Song, et~al.
\newblock The rich-club phenomenon of china's population flow network during the country's spring festival.
\newblock \emph{Applied Geography}, 96:\penalty0 77--85, 2018.

\bibitem[Li and Cai(2007)]{li2007empirical}
W~Li and X~Cai.
\newblock Empirical analysis of a scale-free railway network in china.
\newblock \emph{Physica A: Statistical Mechanics and its Applications}, 382\penalty0 (2):\penalty0 693--703, 2007.

\bibitem[Zhang and Ng(2021)]{zhang2021unveiling}
Yifan Zhang and S~Thomas Ng.
\newblock Unveiling the rich-club phenomenon in urban mobility networks through the spatiotemporal characteristics of passenger flow.
\newblock \emph{Physica A: Statistical Mechanics and its Applications}, 584:\penalty0 126377, 2021.

\bibitem[Zhu et~al.(2021)Zhu, Wang, et~al.]{zhu2021exploring}
Ruoxin Zhu, Yujing Wang, et~al.
\newblock Exploring the rich-club characteristic in internal migration: Evidence from chinese chunyun migration.
\newblock \emph{Cities}, 114:\penalty0 103198, 2021.

\bibitem[Van Den~Heuvel and Sporns(2011)]{van2011rich}
Martijn~P Van Den~Heuvel and Olaf Sporns.
\newblock Rich-club organization of the human connectome.
\newblock \emph{Journal of Neuroscience}, 31\penalty0 (44):\penalty0 15775--15786, 2011.

\bibitem[Ball et~al.(2014)Ball, Aljabar, et~al.]{ball2014rich}
Gareth Ball, Paul Aljabar, et~al.
\newblock Rich-club organization of the newborn human brain.
\newblock \emph{Proceedings of the National Academy of Sciences}, 111\penalty0 (20):\penalty0 7456--7461, 2014.

\bibitem[Jiang and Zhou(2008)]{jiang2008statistical}
Zhi-Qiang Jiang and Wei-Xing Zhou.
\newblock Statistical significance of the rich-club phenomenon in complex networks.
\newblock \emph{New Journal of Physics}, 10\penalty0 (4):\penalty0 043002, 2008.

\bibitem[Berahmand et~al.(2018)Berahmand, Samadi, and Sheikholeslami]{berahmand2018effect}
Kamal Berahmand, Negin Samadi, and Seyed~Mahmood Sheikholeslami.
\newblock Effect of rich-club on diffusion in complex networks.
\newblock \emph{International Journal of Modern Physics B}, 32\penalty0 (12):\penalty0 1850142, 2018.

\bibitem[Cinelli et~al.(2017)Cinelli, Ferraro, and Iovanella]{cinelli2017resilience}
Matteo Cinelli, Giovanna Ferraro, and Antonio Iovanella.
\newblock Resilience of core-periphery networks in the case of rich-club.
\newblock \emph{Complexity}, 2017\penalty0 (1):\penalty0 6548362, 2017.

\bibitem[Colizza et~al.(2006)Colizza, Flammini, et~al.]{colizza2006detecting}
Vittoria Colizza, Alessandro Flammini, et~al.
\newblock Detecting rich-club ordering in complex networks.
\newblock \emph{Nature physics}, 2\penalty0 (2):\penalty0 110--115, 2006.

\bibitem[Hein et~al.(2006)Hein, Schwind, and K{\"o}nig]{hein2006scale}
Oliver Hein, Michael Schwind, and Wolfgang K{\"o}nig.
\newblock Scale-free networks: The impact of fat tailed degree distribution on diffusion and communication processes.
\newblock \emph{Wirtschaftsinformatik}, 48:\penalty0 267--275, 2006.

\bibitem[Barab{\'a}si and Albert(1999)]{barabasi1999emergence}
Albert-L{\'a}szl{\'o} Barab{\'a}si and R{\'e}ka Albert.
\newblock Emergence of scaling in random networks.
\newblock \emph{science}, 286\penalty0 (5439):\penalty0 509--512, 1999.

\bibitem[Derrible and Kennedy(2010)]{derrible2010complexity}
Sybil Derrible and Christopher Kennedy.
\newblock The complexity and robustness of metro networks.
\newblock \emph{Physica A: Statistical Mechanics and its Applications}, 389\penalty0 (17):\penalty0 3678--3691, 2010.

\bibitem[Cohen and Havlin(2010)]{cohen2010complex}
Reuven Cohen and Shlomo Havlin.
\newblock \emph{Complex networks: structure, robustness and function}.
\newblock Cambridge university press, 2010.

\bibitem[Goh et~al.(2001)Goh, Kahng, and Kim]{goh2001universal}
K-I Goh, Byungnam Kahng, and Doochul Kim.
\newblock Universal behavior of load distribution in scale-free networks.
\newblock \emph{Physical review letters}, 87\penalty0 (27):\penalty0 278701, 2001.

\bibitem[Ma and Mondrag{\'o}n(2015)]{ma2015rich}
Athen Ma and Ra{\'u}l~J Mondrag{\'o}n.
\newblock Rich-cores in networks.
\newblock \emph{PloS one}, 10\penalty0 (3):\penalty0 e0119678, 2015.

\bibitem[V{\'a}zquez et~al.(2002)V{\'a}zquez, Pastor-Satorras, and Vespignani]{vazquez2002large}
Alexei V{\'a}zquez, Romualdo Pastor-Satorras, and Alessandro Vespignani.
\newblock Large-scale topological and dynamical properties of the internet.
\newblock \emph{Physical Review E}, 65\penalty0 (6):\penalty0 066130, 2002.

\bibitem[Barth{\'e}lemy(2004)]{barthelemy2004betweenness}
Marc Barth{\'e}lemy.
\newblock Betweenness centrality in large complex networks.
\newblock \emph{The European physical journal B}, 38\penalty0 (2):\penalty0 163--168, 2004.

\bibitem[Masoomy et~al.(2023)Masoomy, Adami, and Najafi]{masoomy2023relation}
H~Masoomy, V~Adami, and MN~Najafi.
\newblock Relation between the degree and betweenness centrality distribution in complex networks.
\newblock \emph{Physical Review E}, 107\penalty0 (4):\penalty0 044303, 2023.

\bibitem[Dorogovtsev et~al.(2006)Dorogovtsev, Goltsev, and Mendes]{dorogovtsev2006k}
Sergey~N Dorogovtsev, Alexander~V Goltsev, and Jose Ferreira~F Mendes.
\newblock K-core organization of complex networks.
\newblock \emph{Physical review letters}, 96\penalty0 (4):\penalty0 040601, 2006.

\bibitem[Newman(2002)]{newman2002assortative}
Mark~EJ Newman.
\newblock Assortative mixing in networks.
\newblock \emph{Physical review letters}, 89\penalty0 (20):\penalty0 208701, 2002.

\bibitem[Krapivsky and Redner(2001)]{krapivsky2001organization}
Paul~L Krapivsky and Sidney Redner.
\newblock Organization of growing random networks.
\newblock \emph{Physical Review E}, 63\penalty0 (6):\penalty0 066123, 2001.

\bibitem[Li et~al.(2011)Li, Zhang, and Small]{li2011emergence}
Ping Li, Jie Zhang, and Michael Small.
\newblock Emergence of scaling and assortative mixing through altruism.
\newblock \emph{Physica A: Statistical Mechanics and its Applications}, 390\penalty0 (11):\penalty0 2192--2197, 2011.

\bibitem[Noldus and Van~Mieghem(2015)]{noldus2015assortativity}
Rogier Noldus and Piet Van~Mieghem.
\newblock Assortativity in complex networks.
\newblock \emph{Journal of Complex Networks}, 3\penalty0 (4):\penalty0 507--542, 2015.

\bibitem[Fotouhi and Rabbat(2013)]{fotouhi2013degree}
Babak Fotouhi and Michael~G Rabbat.
\newblock Degree correlation in scale-free graphs.
\newblock \emph{The European Physical Journal B}, 86:\penalty0 1--19, 2013.

\bibitem[Pek{\"o}z et~al.(2017)Pek{\"o}z, R{\"o}llin, and Ross]{pekoz2017joint}
Erol Pek{\"o}z, Adrian R{\"o}llin, and Nathan Ross.
\newblock Joint degree distributions of preferential attachment random graphs.
\newblock \emph{Advances in Applied Probability}, 49\penalty0 (2):\penalty0 368--387, 2017.

\bibitem[Platt(2019)]{platt2019network}
Edward~L Platt.
\newblock \emph{Network science with Python and NetworkX quick start guide: explore and visualize network data effectively}.
\newblock Packt Publishing Ltd, 2019.

\bibitem[Bayati et~al.(2010)Bayati, Kim, and Saberi]{bayati2010sequential}
Mohsen Bayati, Jeong~Han Kim, and Amin Saberi.
\newblock A sequential algorithm for generating random graphs.
\newblock \emph{Algorithmica}, 58:\penalty0 860--910, 2010.

\bibitem[Cinelli(2019)]{cinelli2019generalized}
Matteo Cinelli.
\newblock Generalized rich-club ordering in networks.
\newblock \emph{Journal of Complex Networks}, 7\penalty0 (5):\penalty0 702--719, 2019.

\end{thebibliography}

\end{document}